\documentclass{INTERSPEECH2023}
\usepackage{tabularx,multirow,cite}
\hypersetup{
    colorlinks=true,
    citecolor=green,
    linkcolor=red,
    urlcolor=magenta,
}
\makeatletter
\newcommand\footnoteref[1]{\protected@xdef\@thefnmark{\ref{#1}}\@footnotemark}
\makeatother

\interspeechcameraready

\title{iSTFTNet2:\\
  Faster and More Lightweight iSTFT-Based Neural Vocoder Using 1D-2D CNN}
\name{Takuhiro Kaneko, Hirokazu Kameoka, Kou Tanaka, Shogo Seki}
\address{
  NTT Communication Science Laboratories, NTT Corporation, Japan}
\email{takuhiro.kaneko@ntt.com}

\begin{document}

\maketitle
 
\begin{abstract}
  The inverse short-time Fourier transform network (iSTFTNet) has garnered attention owing to its fast, lightweight, and high-fidelity speech synthesis. It obtains these characteristics using a fast and lightweight 1D CNN as the backbone and replacing some neural processes with iSTFT. Owing to the difficulty of a 1D CNN to model high-dimensional spectrograms, the frequency dimension is reduced via temporal upsampling. However, this strategy compromises the potential to enhance the speed. Therefore, we propose \textit{iSTFTNet2}, an improved variant of iSTFTNet with a 1D-2D CNN that employs 1D and 2D CNNs to model temporal and spectrogram structures, respectively. We designed a 2D CNN that performs frequency upsampling after conversion in a few-frequency space. This design facilitates the modeling of high-dimensional spectrograms without compromising the speed. The results demonstrated that iSTFTNet2 made iSTFTNet faster and more lightweight with comparable speech quality.\footnote{\label{foot:samples}Audio samples are available at \url{https://www.kecl.ntt.co.jp/people/kaneko.takuhiro/projects/istftnet2/}.}    
\end{abstract}
\noindent\textbf{Index Terms}: speech synthesis, neural vocoder, inverse short-time Fourier transform, convolutional neural network, generative adversarial networks

\section{Introduction}
\label{sec:introduction}

Text-to-speech (TTS) synthesis and voice conversion (VC) have been extensively studied to obtain the desired speech.
The two-stage approach widely used in TTS and VC is as follows.
The first model predicts the intermediate representation (e.g., mel-spectrogram) from the input data (e.g., text or speech), whereas the second model synthesizes speech from the predicted intermediate representation.
This study focuses on the second model, the neural vocoder, and attempts to make it faster and more lightweight to broaden its applicability.

Various neural vocoders have been developed with advances in deep generative models.
The pioneer is an autoregressive model (e.g., WaveNet~\cite{AOordArXiv2016} and WaveRNN~\cite{NKalchbrennerICML2018}) that achieves high-fidelity speech synthesis but suffers from slow inference owing to sample-by-sample processing.
Various parallelizable non-autoregressive models have been developed to boost the inference speed.
For example, successful models include a distillation-based (e.g., Parallel WaveNet~\cite{AOordICML2018} and ClariNet~\cite{WPingICLR2019}), flow (e.g., Glow~\cite{DKingmaNeurIPS2018})-based (e.g., WaveGlow~\cite{RPrengerICASSP2019}), diffusion probabilistic model~\cite{YSongNeurIPS2019,JHoNeurIPS2020}-based (e.g., WaveGrad~\cite{NChenICLR2021} and DiffWave~\cite{ZKongICLR2021}), and generative adversarial network (GAN)~\cite{IGoodfellowNIPS2014}-based (e.g.,~\cite{KKumarNeurIPS2019,RYamamotoICASSP2020,JKongNeurIPS2020,JYangIS2020,GYangSLT2021,AMustafaICASSP2021,JKimIS2021,TOkamotoASRU2021,TKanekoICASSP2022,SHLeeICASSP2022,TKanekoIS2022,YKoizumiSLT2022,TKanekoICASSP2023}) models.
Among them, this study focuses on a GAN-based model while prioritizing the flexibility of the architectural design and the ability of fast inference.

\begin{figure}[t]
  \centering
  \includegraphics[width=0.99\linewidth]{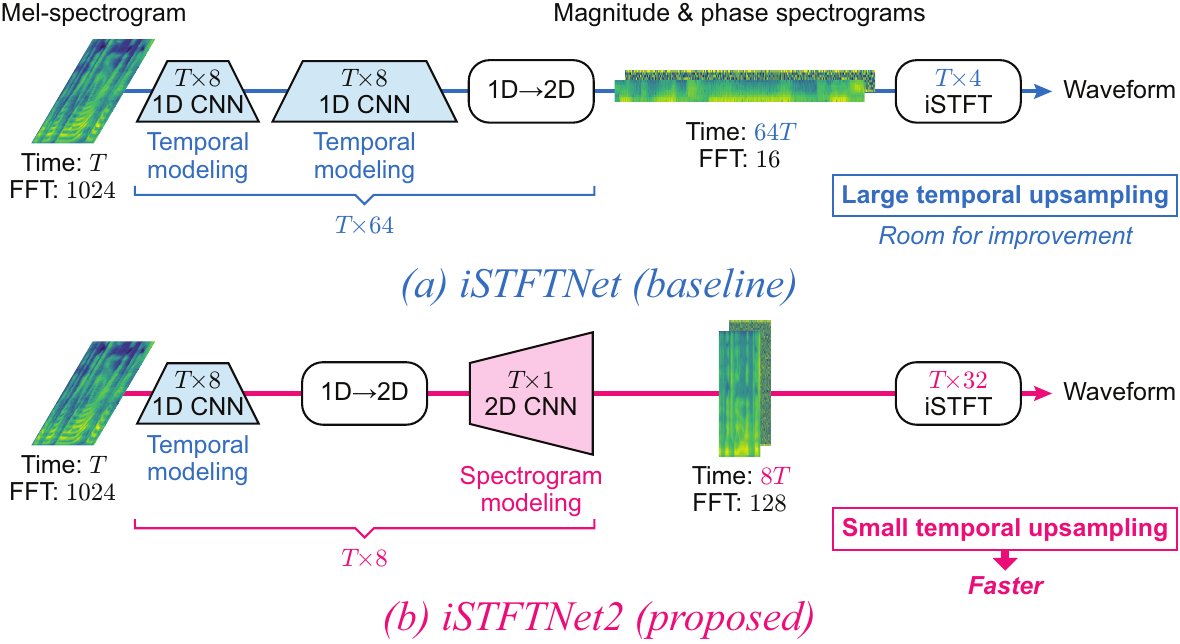}
  \vspace{-1mm}
  \caption{Comparison of iSTFTNet~\cite{TKanekoICASSP2022} and iSTFTNet2 (proposed).
    (a) Owing to the difficulty of a 1D CNN to model high-dimensional spectrograms, it is necessary to reduce the frequency dimension sufficiently in iSTFTNet using large temporal upsampling.
    (b) In contrast, iSTFTNet2 mitigates this difficulty by conducting 1D-to-2D conversion in an earlier stage and applying a 2D CNN that can capture local structures in spectrograms.
    This modification facilitates the reduction of the neural temporal upsampling by eight times (i.e., from $\times 64$ to $\times 8$) and enhances the inference speed.}
  \label{fig:concept}
  \vspace{-3mm}
\end{figure}

Among the GAN-based neural vocoders, one of the fastest and most lightweight models is the inverse short-time Fourier transform network (iSTFTNet)~\cite{TKanekoICASSP2022}, which achieves fast, lightweight, and high-fidelity speech synthesis using a fast and lightweight 1D CNN (e.g., HiFi-GAN~\cite{JKongNeurIPS2020}) as the backbone and replacing some output-side neural processes with fast and lightweight inverse short-time Fourier transform (iSTFT).
Particularly, iSTFTNet applies iSTFT after sufficiently reducing the frequency dimension using large temporal upsampling (Figure~\ref{fig:concept}(a)) to avoid modeling high-dimensional spectrograms, which are difficult for a 1D CNN to represent.
This technique is essential for making the model faster and more lightweight while maintaining speech quality; however, it compromises the potential to improve the inference speed by applying iSTFT with fewer temporal upsampling.

\begin{figure*}[t]
  \centering
  \includegraphics[width=0.99\textwidth]{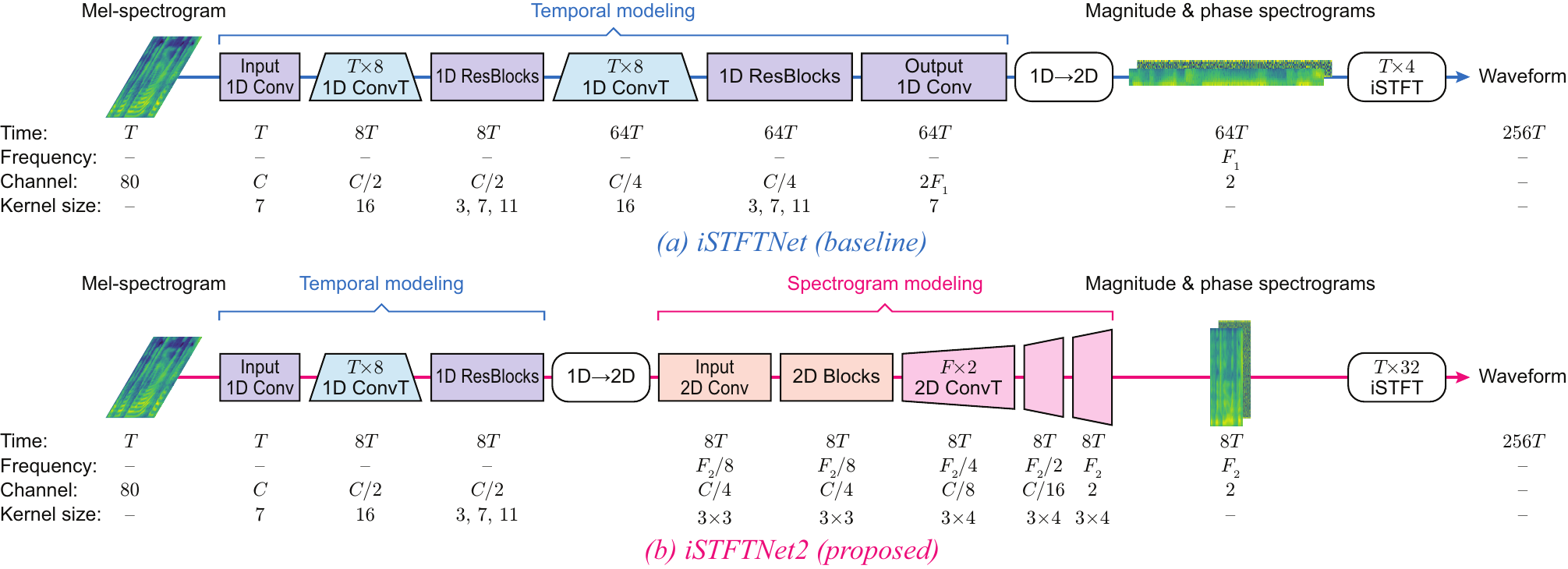}
  \vspace{-2mm}
  \caption{Overall architectures of iSTFTNet~\cite{TKanekoICASSP2022} and iSTFTNet2 (proposed) when incorporated into HiFi-GAN~\cite{JKongNeurIPS2020}.
    $T$ and $C$ denote the temporal length and number of output channels, respectively.
    $F_1$ indicates the frequency dimension of the spectrogram in iSTFTNet, whereas $F_2$ indicates that in iSTFTNet2.
    $C$ is set to $128$ when HiFi-GAN V2 (lightweight variant) is used as the backbone.
    $F_1$ and $F_2$ can be calculated by $\frac{f_s}{2} + 1 = \frac{f_1 / s}{2} + 1$ (see Equation~\ref{eq:istft} for the definition of $f_s$).
    When the FFT size ($f_1$) is $1024$, they are calculated as $F_1 = \frac{1024 / 64}{2} + 1 = 9$ and $F_1 = \frac{1024 / 8}{2} + 1 = 65$.}
  \label{fig:architecture}
  \vspace{-3mm}
\end{figure*}

One possible solution is to conduct spectrogram conversion using a fully 2D CNN (e.g.,~\cite{TKanekoIS2017,KOyamadaEUSIPCO2018,PNeekharaIS2019}).
However, the direct application of a 2D CNN requires a significant increase in the calculation cost because it increases linearly according to the frequency dimension (e.g., $80$ in a mel-spectrogram).
Alternatively, inspired by the success of combining 1D and 2D CNNs~\cite{TKanekoICASSP2019}, we propose \textit{iSTFTNet2}, an improved variant of iSTFTNet with a 1D-2D CNN, in which 1D and 2D CNNs are used to model the global temporal and local spectrogram structures, respectively (Figure~\ref{fig:concept}(b)).
Particularly, we designed a 2D CNN that conducts frequency upsampling after performing sufficient conversion in a few-frequency space using 1D and few-frequency 2D CNNs.
This design facilitates the modeling of high-dimensional spectrograms, which are difficult for a conventional 1D CNN-based iSTFTNet to model, without compromising the speed.
Furthermore, we propose an efficient module inspired by ShuffleNets~\cite{XZhangCVPR2018,NMaECCV2018} to improve the speed and model size further.

In the experiments, we examined the effectiveness of iSTFTNet2 on two representative datasets: \textit{LJSpeech}~\cite{ljspeech17} (single English speaker) and \textit{VCTK}~\cite{JYamagishiCSTR2016} (multiple English speakers).
The experimental results demonstrated that iSTFTNet2 made iSTFTNet faster and more lightweight with comparable speech quality.
Furthermore, we demonstrated the versatility of our ideas by applying iSTFTNet2 to multi-band modeling~\cite{CYuIS2020,GYangSLT2021}, another technique for improving the speed.
The results showed that this variant could further improve the speed with comparable speech quality.

The remainder of this paper is organized as follows.
Section~\ref{sec:istftnet} briefly reviews the conventional iSTFTNet.
Section~\ref{sec:istftnet2} presents details of the proposed iSTFTNet2.
Section~\ref{sec:experiments} presents our experimental results.
Finally, Section~\ref{sec:conclusions} concludes the study and discusses future research.

\section{Preliminary: Conventional iSTFTNet}
\label{sec:istftnet}

iSTFTNet~\cite{TKanekoICASSP2022} is one of the fastest and most lightweight neural vocoders.
These characteristics were obtained by replacing some of the output-side layers of a fully neural vocoder with fast and lightweight iSTFT.
Particularly, it uses a fast and lightweight 1D CNN (e.g., HiFi-GAN~\cite{JKongNeurIPS2020}) as the backbone with a high processing speed.
However, it is challenging for a 1D CNN to model high-dimensional spectrograms because of the difficulty in capturing local structures in the frequency direction.
Hence, iSTFTNet reduces the frequency dimension using temporal upsampling as follows:
\begin{flalign}
  \label{eq:istft}
  \text{iSTFT} (f_s, h_s, w_s)
  = \text{iSTFT} \left( \frac{f_1}{s}, \frac{h_1}{s}, \frac{w_1}{s} \right),
\end{flalign}
where $f_s$, $h_s$, and $w_s$ indicate the FFT size, hop length, and window length, respectively, required for the iSTFT after $\times s$ temporal upsampling.
This equation is based on the time and frequency tradeoff, that is, $f_1 \cdot 1 = f_s \cdot s = \text{constant}$, and indicates that the frequency dimension can be reduced $s$ times by conducting $\times s$ temporal upsampling.

Figure~\ref{fig:architecture}(a) shows the overall architecture of iSTFTNet.
We present the architecture of iSTFTNet-\texttt{C8C8I4},\footnote{This is the same as iSTFTNet-\texttt{C8C8} described in~\cite{TKanekoICASSP2022}.
  Here, we added \texttt{I4} to specify the temporal upsampling scale in the iSTFT.} which is the best balanced model that improves the speed and model size while maintaining speech quality, where \texttt{C}$x$ indicates the use of 1D CNN blocks with $\times x$ temporal upsampling, and \texttt{I}$y$ indicates the use of iSTFT with $\times y$ temporal upsampling.
When prioritizing the speed, a model that performs temporal upsampling fewer times, for example, iSTFTNet-\texttt{C8C1I32}, which conducts temporal upsampling once, is better.
However, it is shown that such a model deteriorates speech quality because of the difficulty of a 1D CNN in modeling high-dimensional spectrograms~\cite{TKanekoICASSP2022}.

\section{Proposal: iSTFTNet2}
\label{sec:istftnet2}

Considering the abovementioned facts, we attempted to construct an improved variant of iSTFTNet that can maintain speech quality even with fewer temporal upsampling.
A possible solution is to convert a spectrogram using a fully 2D CNN that can capture the local structures in spectrograms (e.g.,~\cite{TKanekoIS2017,KOyamadaEUSIPCO2018,PNeekharaIS2019}).
However, this replacement requires a significant increase in the calculation cost because it increases linearly in proportion to the frequency dimension (e.g., $80$ in a mel-spectrogram).

Alternatively, we developed \textit{iSTFTNet2}, which constitutes a 1D-2D CNN.
Figure~\ref{fig:architecture}(b) presents the overall architecture of iSTFTNet2.
As shown in this figure, to model temporal structures efficiently, iSTFTNet2 uses the same 1D CNN for the first three modules as that used in iSTFTNet, except that channel concatenation is used instead of addition when integrating the outputs of the multi-receptive fusion~\cite{JKongNeurIPS2020} in the 1D ResBlock to propagate more information to the subsequent 2D CNN.
Unlike iSTFTNet, iSTFTNet2 conducts 1D-to-2D conversion in an earlier step and applies a 2D CNN to effectively capture the local structures in the spectrograms, which are difficult for a 1D CNN to model.

When considering the detailed configuration of a 2D CNN, it is important to prevent an increase in the calculation cost driven by the introduction of a 2D CNN because its calculation cost increases linearly in proportion to the time and frequency dimensions.
To address this problem, iSTFTNet2 performs the main conversion in a few-frequency space (specifically, 2D blocks are applied in a space in which the frequency dimension is downsampled eight times, as shown in Figure~\ref{fig:architecture}(b)), and then conducts frequency upsampling in the last phase using transposed convolutions.

\begin{figure}[t]
  \centering
  \includegraphics[width=0.99\linewidth]{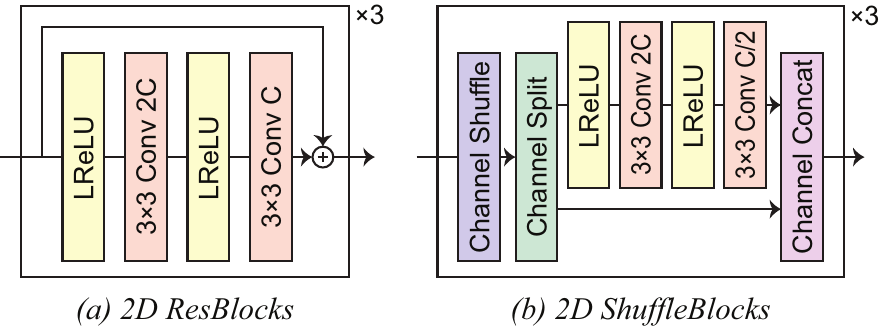}
  \vspace{-1mm}
  \caption{Architectures of 2D blocks used in iSTFTNet2.
    ``LReLU'' denotes a leaky rectified linear unit~\cite{AMaasICML2013} with a negative slope of $0.1$.
    ``$x \times y$ Conv $z$'' indicates a 2D convolution with a kernel size of $x \times y$ and the number of output channels of $z$, and $C$ in this module denotes the number of input channels of the block.
    In (a), residual connection (which is denoted by ``+'')~\cite{KHeCVPR2016} is used.
    In (b), the weight-free operations used in ShuffleNets~\cite{XZhangCVPR2018,NMaECCV2018}, i.e., ``Channel Shuffle,'' ``Channel Split,'' and ``Channel Concat,'' are used.
    In both blocks, the presented block is stacked three times (which is denoted by ``$\times 3$'').}
  \label{fig:blocks}
  \vspace{-3mm}
\end{figure}

The design of the 2D blocks is a vital aspect to consider.
As shown in Figure~\ref{fig:blocks}, we developed two architectural designs.
The first is a \textit{2D ResBlock} (Figure~\ref{fig:blocks}(a)) that uses a residual connection~\cite{KHeCVPR2016} to propagate information efficiently.
We adjusted the model parameters (i.e., number of channels and kernel size) such that the model became faster and more lightweight than iSTFTNet-\texttt{C8C8I4} (the best-balanced model).
To further make the model faster and more lightweight, we introduced a second model, that is, a \textit{2D ShuffleBlock} (Figure~\ref{fig:blocks}(b)), which is inspired by efficient neural networks called ShuffleNets~\cite{XZhangCVPR2018,NMaECCV2018}.
In this block, the number of model parameters used in the 2D convolutional layers was adjusted such that it was half of that of the 2D ResBlock (Figure~\ref{fig:blocks}(a)).
Alternatively, in contrast to the residual connection, the half channels are propagated directly without any addition to preserve the model capacity.
A channel shuffle~\cite{XZhangCVPR2018,NMaECCV2018} is conducted to provide an interaction between the skip and non-skip branches.
Because the channel shuffle, channel split, and channel concat are weight-free operations, this block is faster and more lightweight than the 2D ResBlock.
We demonstrated the empirical performance difference between the two 2D blocks in the experiments presented in the next section.

\section{Experiments}
\label{sec:experiments}

\subsection{Experimental setup}
\label{subsec:experimental_setup}

\textbf{Dataset.}
We examined the effectiveness of iSTFTNet2 using two representative datasets.
\textit{LJSpeech}~\cite{ljspeech17}, which includes 13,100 audio clips of a single English speaker, and 12,500, 100, and 500 audio clips were used for training, validation, and evaluation, respectively. 
\textit{VCTK}~\cite{JYamagishiCSTR2016}, which comprises 44,081 audio clips of 108 different English speakers, and 41,921, 1,080, and 1,080 audio clips were  used for training, validation, and evaluation, respectively.
Following a study on HiFi-GAN~\cite{JKongNeurIPS2020}, audio clips were sampled at $22.05$ kHz, and $80$-dimensional log-mel spectrograms were extracted from the audio clips with an FFT size of $1024$, a hop length of $256$, and a window length of $1024$.

\smallskip\noindent\textbf{Implementation.}
We used \textit{iSTFTNet-\texttt{C8C8I4}}~\cite{TKanekoICASSP2022} as a baseline because it is the best-balanced model and its speech quality has been demonstrated~\cite{TKanekoICASSP2022} to be comparable to that of HiFi-GAN~\cite{JKongNeurIPS2020}, which is a widely used baseline model.
Specifically, we implemented iSTFTNet-\texttt{C8C8I4} based on \textit{HiFi-GAN V2} (a lightweight variant) because we were interested in the performance of the lightweight model.\footnote{In additional experiments, we also examined the performance when \textit{HiFi-GAN V1} (a high-quality variant) was used as a baseline, and observed a similar tendency in terms of cFW2VD, RTF, and \#~Param.}
We implemented iSTFTNet2 by replacing the modules of iSTFTNet-\texttt{C8C8I4}, as shown in Figure~\ref{fig:architecture}.
Particularly, we implemented the two variants using different 2D blocks, as shown in Figure~\ref{fig:blocks}.
For clarity, we denote iSTFTNet2 with 2D ResBlocks (Figure~\ref{fig:blocks}(a)) and that with 2D ShuffleBlocks (Figure~\ref{fig:blocks}(b)) as \textit{iSTFTNet2-Base} and \textit{iSTFTNet2-Small}, respectively.
As another comparison model, we examined \textit{iSTFTNet-\texttt{C8C1I32}}, which conducts the same temporal upsampling as iSTFTNet2 but uses 1D ResBlocks instead of 2D blocks unlike iSTFTNet2.
This model was used to validate the importance of 2D blocks.
All models were implemented based on open-source code,\footnote{\label{foot:parallel_wavegan}\url{https://github.com/kan-bayashi/ParallelWaveGAN}} and the same training settings were used.
Specifically, a combination of least-squares GAN~\cite{XMaoICCV2017}, mel-spectrogram~\cite{JKongNeurIPS2020}, and feature matching ~\cite{ALarsenICML2016,KKumarNeurIPS2019} losses was used as the loss function.
Each model was trained for $2.5$M iterations using the Adam optimizer~\cite{DPKingmaICLR2015} with a batch size of $16$, an initial learning rate of $0.0002$, and momentum terms $\beta_1$ and $\beta_2$ of $0.5$ and $0.9$, respectively.

\smallskip\noindent\textbf{Evaluation metrics.}
We conducted mean opinion score (\textit{MOS}) tests to evaluate perceptual quality.
Twenty audio clips were randomly selected from the evaluation set, and log-mel spectrograms extracted from the audio clips were used as vocoder inputs.
In addition to the speech synthesized by the abovementioned models, \textit{ground-truth} speech was included as an anchor sample.
Ten listeners participated in each online test and were asked to assess speech quality using a five-grade evaluation: 1 = bad, 2 = poor, 3 = fair, 4 = good, and 5 = excellent.
As an objective metric, we used the conditional Fr\'{e}chet wav2vec distance (\textit{cFW2VD})~\cite{TKanekoICASSP2022}, which measures the distribution distance between the real and synthesized speech in a wav2vec 2.0~\cite{ABaevskiNeurIPS2020} feature space conditioned on the text.
The smaller the value, the higher the similarity.
We evaluated the inference speed using a real-time factor (\textit{RTF}) that was calculated by dividing the inference time by the duration of the synthesized speech (fixed at $1$~s in the experiments).
The RTF was measured using a single thread on an Intel Core i7-12700H CPU.
The smaller the value, the higher the speed.
We evaluated the model size using the number of parameters (\textit{\#~Param}).
The smaller the value, the more lightweight the model.
The audio samples are available from the link on the first page.\footnoteref{foot:samples}

\subsection{Results on single speaker dataset}
\label{subsec:results_ljspeech}

Table~\ref{tab:results_ljspeech} summarizes the results on LJSpeech.
These results are discussed from three perspectives.

\smallskip\noindent\textbf{Speech quality.}
For the MOS test, we conducted the Mann--Whitney U test.
We found that iSTFTNet-\texttt{C8C8I4}, iSTFTNet2-Base, and iSTFTNet2-Small were \textit{not} significantly different from HiFi-GAN V2 in terms of the $p$-values $> 0.05$.
In contrast, iSTFTNet-\texttt{C8C8I32} performed significantly worse than the others.
cFW2VD was also the worst in iSTFTNet-\texttt{C8C8I32} and was comparable in the other cases.
These results indicate that iSTFTNet2 can be used as an alternative to iSTFTNet and HiFi-GAN regarding speech quality.

\smallskip\noindent\textbf{Inference speed.}
The RTF shows that both iSTFTNet2-Base and iSTFTNet2-Small were faster than HiFi-GAN and iSTFTNet-\texttt{C8C8I4}, achieving comparable speech quality.
iSTFTNet2-Small was the fastest among them and was comparable to iSTFTNet-\texttt{C8C8I32}, which sacrifices speech quality.

\smallskip\noindent\textbf{Model size.}
We found that iSTFTNet2-Base and iSTFTNet2-Small were lighter than all baselines, and iSTFTNet2-Small was the most lightweight in terms of \# Param.\footnote{\# Param of iSTFTNet-\texttt{C8C1I32} is larger than that of iSTFTNet-\texttt{C8C8I4} because the number of channels is halved in the second 1D ResBlocks in iSTFTNet-\texttt{C8C8I4}, whereas it is not conducted in iSTFTNet-\texttt{C8C1I32} owing to the absence of temporal upsampling.
  We used this strategy to confirm whether iSTFTNet-\texttt{C8C1I32} could not obtain comparable speech quality, even with expressive modules.}

\begin{table}[t]
  \caption{Comparison of MOS with 95\% confidence intervals, cFW2VD, RTF, and \#~Param on LJSpeech.
    The numbers in () indicate the rates (\%) compared with HiFi-GAN V2.}
  \vspace{-2mm}
  \label{tab:results_ljspeech}
  \newcommand{\spm}[1]{{\tiny$\pm$#1}}
  \newcommand{\rate}[1]{{\tiny({#1})}}
  \setlength{\tabcolsep}{4.5pt}
  \centering
  \scriptsize{
    \begin{tabularx}{\columnwidth}{lcccc}
      \toprule
      \multicolumn{1}{c}{\textbf{Model}} & \textbf{MOS$\uparrow$} & \textbf{cFW2VD$\downarrow$} & \textbf{RTF$\downarrow$} & \textbf{\#~Param$\downarrow$}
      \\ \midrule
      Ground truth
      & 4.71 \spm{0.07} & -- & -- & --
      \\ \midrule
      HiFi-GAN V2
      & 4.20 \spm{0.10} & 0.046 & 0.053 \rate{100}                & 0.93M \rate{100}
      \\ \midrule
      iSTFTNet-\texttt{C8C8I4}
      & 4.12 \spm{0.10} & 0.042 & 0.029 \rate{\hspace{0.25em} 55} & 0.89M \rate{\hspace{0.25em} 96}
      \\
      iSTFTNet-\texttt{C8C1I32}
      & 3.71 \spm{0.13} & 0.071 & 0.018 \rate{\hspace{0.25em} 34} & 1.30M \rate{140}
      \\ \midrule
      iSTFTNet2-Base
      & 4.24 \spm{0.10} & 0.036 & 0.021 \rate{\hspace{0.25em} 41} & 0.85M \rate{\hspace{0.25em} 91}
      \\
      iSTFTNet2-Small
      & 4.22 \spm{0.10} & 0.040 & 0.018 \rate{\hspace{0.25em} 35} & 0.79M \rate{\hspace{0.25em} 85}
      \\ \bottomrule
    \end{tabularx}
  }
\end{table}

\subsection{Results on multiple speaker dataset}
\label{subsec:results_vctk}

Table~\ref{tab:results_vctk} lists the results on VCTK.
We observed that the same tendency as that observed on LJSpeech.
For the MOS test, iSTFTNet-\texttt{C8C8I4}, iSTFTNet2-Base, and iSTFTNet2-Small were \textit{not} significantly different from HiFi-GAN V2 in terms of the $p$-values $> 0.05$ in the Mann--Whitney U test, whereas iSTFTNet-\texttt{C8C8I32} performed significantly worse than the others.
The RTF and \#~Param were the same as those observed on LJSpeech.

\begin{table}[t]
  \caption{Comparison of MOS with 95\% confidence intervals, cFW2VD, RTF, and \#~Param on VCTK.
    The numbers in () indicate the rates (\%) compared with HiFi-GAN V2.}
  \vspace{-2mm}
  \label{tab:results_vctk}
  \newcommand{\spm}[1]{{\tiny$\pm$#1}}
  \newcommand{\rate}[1]{{\tiny({#1})}}
  \setlength{\tabcolsep}{4.5pt}
  \centering
  \scriptsize{
    \begin{tabularx}{\columnwidth}{lcccc}
      \toprule
      \multicolumn{1}{c}{\textbf{Model}} & \textbf{MOS$\uparrow$} & \textbf{cFW2VD$\downarrow$} & \textbf{RTF$\downarrow$} & \textbf{\#~Param$\downarrow$}
      \\ \midrule
      Ground truth
      & 4.38 \spm{0.09} & -- & -- & --
      \\ \midrule
      HiFi-GAN V2
      & 3.99 \spm{0.11} & 0.061 & 0.053 \rate{100}                & 0.93M \rate{100}
      \\ \midrule
      iSTFTNet-\texttt{C8C8I4}
      & 3.94 \spm{0.12} & 0.065 & 0.029 \rate{\hspace{0.25em} 55} & 0.89M \rate{\hspace{0.25em} 96}
      \\
      iSTFTNet-\texttt{C8C1I32}
      & 3.40 \spm{0.13} & 0.110 & 0.018 \rate{\hspace{0.25em} 34} & 1.30M \rate{140}
      \\ \midrule
      iSTFTNet2-Base
      & 3.91 \spm{0.11} & 0.062 & 0.021 \rate{\hspace{0.25em} 41} & 0.85M \rate{\hspace{0.25em} 91}
      \\
      iSTFTNet2-Small
      & 3.91 \spm{0.12} & 0.067 & 0.018 \rate{\hspace{0.25em} 35} & 0.79M \rate{\hspace{0.25em} 85}
      \\ \bottomrule
    \end{tabularx}
  }
  \vspace{-1mm}
\end{table}

\subsection{Application to multi-band modeling}
\label{subesc:application}

As discussed in~\cite{TKanekoICASSP2022}, iSTFT and multi-band modeling~\cite{CYuIS2020,GYangSLT2021} (another technique for improving speed) are complementary, and the speed can be further enhanced by combining them using $\text{iSTFT}\left( \frac{f_1}{sb}, \frac{h_1}{sb}, \frac{w_1}{sb} \right)$, where $b$ is the number of sub-bands.
Motivated by this fact, we evaluated the performance of applying iSTFTNet2 to multi-band modeling.
We examined the effectiveness of this variant on LJSpeech~\cite{ljspeech17}.

\smallskip\noindent\textbf{Implementation.}
As a baseline, we used iSTFTNet-\texttt{C4C4I4B4}, where \texttt{C}/\texttt{I}/\texttt{B}$x$ indicates the use of 1D blocks/iSTFT/multi-band modeling with $\times x$ temporal upsampling.
We denote this model as \textit{iSTFTNet-MB}.
We modified this model to iSTFTNet2 by replacing the second \texttt{C4} with 2D ShuffleBlocks (Figure~\ref{fig:blocks}(b)) and using \texttt{I16} instead of \texttt{I4}.
The number of final output channels of the 2D CNN ($2$ in Figure~\ref{fig:architecture}) was modified to $8$ to produce four sub-bands.
To allow for this expansion, we doubled the number of channels in the 2D CNN and alternatively changed the number of output channels in the first convolution layer in a 2D ShuffleBlock (Figure~\ref{fig:blocks}(b)) from $2C$ to $C$ to make the model size and inference speed similar to those of iSTFTNet-MB.
We denote this model as \textit{iSTFTNet2-MB}.

\smallskip\noindent\textbf{Results.}
Table~\ref{tab:results_application} lists the results.
The RTF and \#~Param were almost the same for iSTFTNet-MB and iSTFTNet2-MB because we adjusted the model parameters of iSTFTNet2-MB such that they were almost the same.
However, iSTFTNet2-MB significantly outperformed iSTFTNet-MB in terms of MOS (with the $p$-value $< 0.05$ in the Mann--Whitney U test) and cFW2VD.
This is possibly because it is easier to represent multiple sub-band spectrograms simultaneously in a 2D CNN (in which channels and frequencies are represented in independent dimensions) than in a 1D CNN (in which they are mixed in the same dimension).
Furthermore, iSTFTNet2-MB was \textit{not} significantly different from HiFi-GAN V2 (Table~\ref{tab:results_ljspeech}) in these metrics (for the MOS, the $p$-value $> 0.05$ in the Mann--Whitney U test), while reducing the RTF to $21\%$.
These results indicated that iSTFTNet2-MB was the best among the variants of iSTFTNets and iSTFTNet2s when prioritizing speed and speech quality.

\begin{table}[t]
  \caption{Comparison of MOS with 95\% confidence intervals, cFW2VD, RTF, and \#~Param when incorporating multi-band modeling on LJSpeech.
    The numbers in () indicate the rates (\%) compared with HiFi-GAN V2.}
  \vspace{-2mm}
  \label{tab:results_application}
  \newcommand{\spm}[1]{{\tiny$\pm$#1}}
  \newcommand{\rate}[1]{{\tiny({#1})}}
  \setlength{\tabcolsep}{6pt}
  \centering
  \scriptsize{
    \begin{tabularx}{\columnwidth}{lcccc}
      \toprule
      \multicolumn{1}{c}{\textbf{Model}} & \textbf{MOS$\uparrow$} & \textbf{cFW2VD$\downarrow$} & \textbf{RTF$\downarrow$} & \textbf{\#~Param$\downarrow$}
      \\ \midrule
      Ground truth
      & 4.71 \spm{0.07} & -- & -- & --
      \\ \midrule
      iSTFTNet-MB
      & 4.05 \spm{0.12} & 0.061 & 0.012 \rate{\hspace{0.25em} 22} & 0.82M \rate{\hspace{0.25em} 88}
      \\ \midrule
      iSTFTNet2-MB
      & 4.25 \spm{0.11} & 0.040 & 0.011 \rate{\hspace{0.25em} 21} & 0.83M \rate{\hspace{0.25em} 89}
      \\ \bottomrule
    \end{tabularx}
  }
  \vspace{-1mm}
\end{table}

\section{Conclusions}
\label{sec:conclusions}

We proposed \textit{iSTFTNet2}, an improved variant of iSTFTNet that constitutes a 1D-2D CNN, in which 1D and 2D CNNs are used to model temporal and spectrogram structures, respectively.
The proposed architecture facilitated the application of iSTFT to higher-dimensional spectrograms without large temporal upsampling, and the experimental results demonstrated that iSTFTNet2 made iSTFTNet faster and more lightweight while maintaining speech quality.
Although we focused on a GAN-based neural vocoder, our ideas have high applicability, and applying them to other models, including other neural vocoders (e.g.,~\cite{RPrengerICASSP2019,NChenICLR2021,ZKongICLR2021}) and end-to-end text-to-speech synthesis (e.g.,~\cite{WPingICLR2019,YRenICLR2021,JDonahueICLR2021,JKimICML2021,NChenIS2021,XTanArXiv2022}), remains the subject of future research.

\section{Acknowledgements}
This work was supported by JST CREST Grant Number JPMJCR19A3, Japan.

\clearpage
\bibliographystyle{IEEEtran}
\bibliography{refs}

\end{document}